\documentclass[12pt,twoside]{article}
\usepackage{amsmath}

\input BoxedEPS
\SetRokickiEPSFSpecial  %% for dvips by Tom Rokicki, for VMS
\HideDisplacementBoxes

\raggedright

\newcommand{\rd}[1]{\mathop{\mathrm{d}#1}}

\newcommand{\grad}{\vec\nabla}

\newcommand{\Loch}{Loch\-lainn}
\newcommand{\Oraf}{O'Raif\-ear\-taigh}

\let\ts\textsuperscript

\newcommand{\numeq}[2]{\begin{equation}
#2
\label{#1}
\end{equation}}
\newcommand{\refeq}[1]{(\ref{#1})}

\let\vec\boldsymbol
\let\eps\varepsilon
\let\epsilon\varepsilon
\let\phi\varphi

\begin{document}
 
\title{\Loch\ \Oraf,\\ Fluids,  and Noncommuting Fields}
\author{R. Jackiw\\
\small\it Center for Theoretical Physics\\ 
\small\it Massachusetts Institute of Technology\\ 
\small\it Cambridge, MA 02139-4307
}%\\[1ex]
%\small\sl(American Association of  Physics Teachers,   Philadelphia, January 2002)}
\date{}
%\\

%\date{\small Typeset in \LaTeX\ by M. Stock\\
%MIT-CTP\#3043}
\maketitle

\abstract{\noindent
\Loch\ \Oraf\ and his work are recalled; the connection between fluid mechanics -- his last
research topic --  and noncommuting gauge fields is explained.}

\vfill

\centerline{\large \Oraf\ Memorial Meeting, Dublin, Ireland, September 2002}

\pagestyle{myheadings}
\markboth{\small {\it R. Jackiw}}{\small  \Loch\ \Oraf, Fluids,  and Noncommuting Fields}
\thispagestyle{empty}

\newpage

\null\vspace*{1pc}

\centerline{\BoxedEPSF{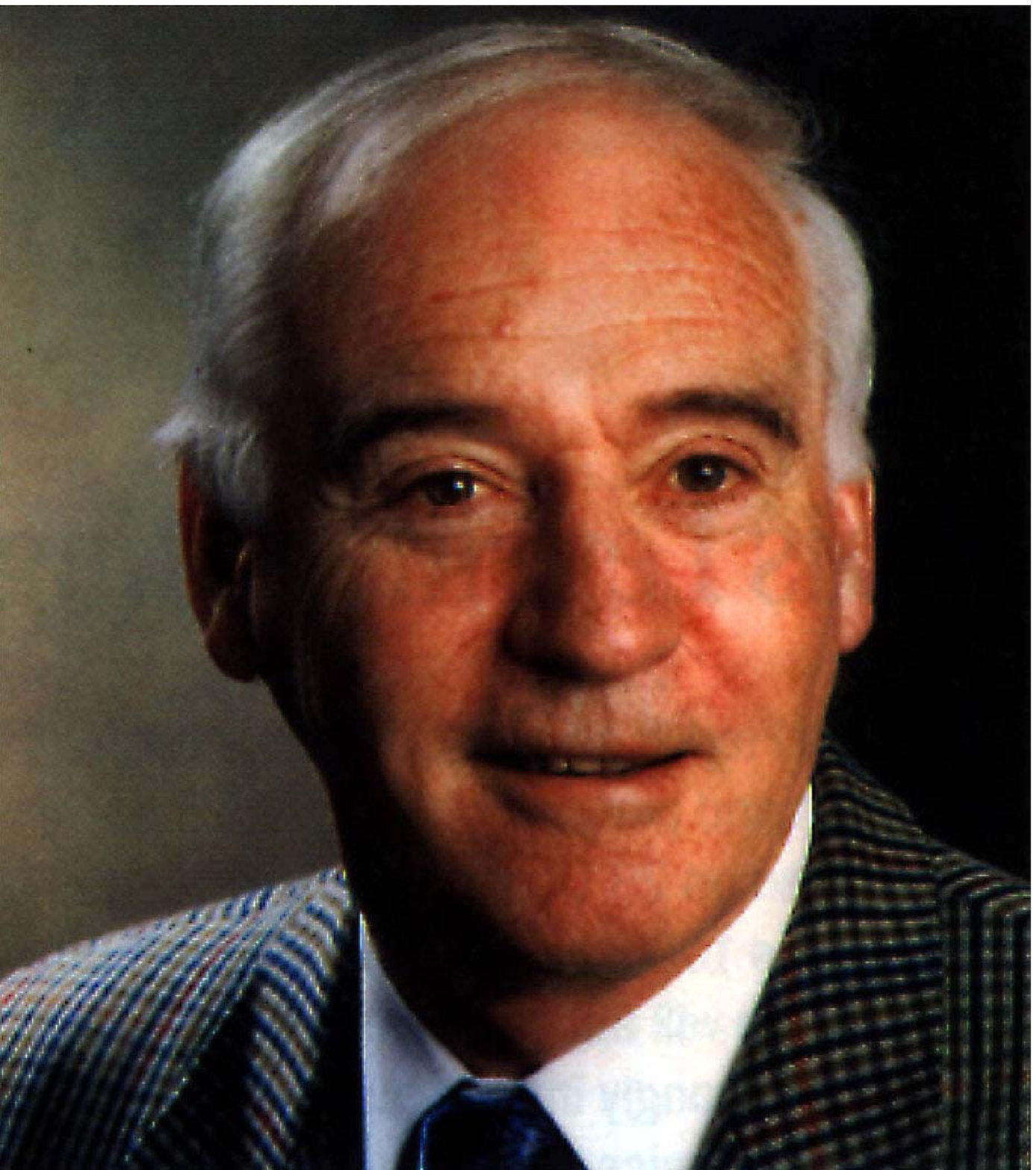 scaled 900}}

\thispagestyle{empty}
\newpage

\vspace*{-.5truein}

\section*{}

\Loch\ \Oraf\ was an eminent theoretical/mathematical physicist, the most recent member of a
cohort of outstanding Irish physicists that includes Hamilton, Stokes, Father Callan (the
inventor of the induction coil), Fitzgerald, Walton, Synge, and others. I came to know him in the
1970s during his  visits to  my colleague the mathematician Irving Segal. Of course, I already
knew his famous work from the 1960s, wherein he put a stop to a search for a
group-theoretical combination of internal and kinematical relativistic symmetries. At that
time, particle physicists were impressed by Wigner's successful combination of spin SU(2) and
isospin SU(2) into an SU(4) of nuclear physics, and they tried to find something similar for the
quark model. The obvious choice was an SU(6) built from spin and Gell-Mann's  flavor SU(3),
but this was a nonrelativistic construction and the search was on for a relativistic version.
Various authors pointed out various difficulties with the idea, but the definitive paper was
\Loch's \emph{Physical Review Letter}$^1$ in which he proved that unacceptable mass
degeneracies follow when the putative invariance group contains internal symmetries and
Poincar\'e symmetries, mixed in a nontrivial way. This story is documented in Dyson's
collection \emph{Symmetry Groups in Nuclear and Particle Physics}.$^2$ 

Although more definitive no-go theorems were later constructed,$^3$ serious research on
``relativistic SU(6)'' stopped with the publication of ``\Oraf's theorem''.  But the desire of
physicists to combine Poincar\'e and internal symmetries did not stop, and this is an
interesting example of how the force of physics, which derives from Nature, overcomes
mathematical obstacles, which are put up by human ingenuity! Of course, I am referring to
supersymmetry, which evades all no-go theorems and does combine internal with space-time
symmetries  through the simple expedient of mixing bosons and fermions -- something that
does not happen for transformations belonging to ordinary symmetry groups. Again, \Loch\
contributed decisively to this newly evolved idea. Supersymmetry still entails unacceptable
mass degeneracies, between bosons and fermions. In his most-cited paper on the ``\Oraf\
mechanism'', he constructed simple examples in which supersymmetry is spontaneously
broken, and the unwanted  mass-degeneracies are removed.\ts4

In later years \Loch\ wrote extensively on historical and pedagogical topics, centered on gauge
theories. I recommend especially his two books: \emph{Group Structure of Gauge
Theories}$^5$ and
\emph{The Dawning of Gauge Theory},$^6$ and two articles: ``Gauge Theory: The Gentle
Revolution''$^7$ and ``Gauge Theory: Historical Origins and Some Modern Developments'',$^8$
the last with Straumann.   These works reflect his life-long desire to teach, clarify, and explain
physics intricacies, and this is also seen from the vast number of schools, symposia, and
meetings to  which he contributed -- I counted twenty-five in  the last decade. Both the
research and the teaching were fittingly recognized by the Wigner medal, which he received
just before he died. 

\Loch\ and I never collaborated on actual research, but our interests paralleled each other to
a significant degree. Both he and I explored anomalies; we both wrote on effective
potentials; he worked extensively on monopoles, and I determined the quantum-mechanical
implications of these and other classical solutions. His last published paper with Sreedhar$^9$
concerns fluid mechanics, and I too have recently been studying this topic because of its
unexpected connections to extended objects in field theory and to noncommutative gauge
theories.\ts{10} \Loch\ did not have the opportunity to work on noncommuting
gauge theories, but I suspect that he would have liked to, because that subject fits so well with
everything that he did before. So I shall conclude my talk by informing you about the
relevance of fluids to noncommuting gauge fields. 

The suggestion that configuration-space coordinates may not commute
\numeq{e1}{
[x^i, x^j] = i\theta^{ij}
}
where $\theta^{ij}$ is a constant, anti-symmetric two-index object, has arisen recently from
string theory, but in fact it has a longer history.  Like many interesting quantum-mechanical
ideas, it was first suggested  by Heisenberg, in the late 1930s, who reasoned that coordinate
noncommutativity would entail a coordinate uncertainty and would ameliorate short-distance
singularities, which beset quantum fields. He told  his idea to Peierls, who eventually made use
of it when analyzing electronic systems in an external magnetic field, so strong that projection
to the lowest Landau level is justified. After this projection, the coordinates fail to commute
(since the state space has been truncated).\ts{11}  But this phenomenological realization of
Heisenberg's idea did not address  issues in fundamental science, so Peierls told Pauli about
it, who in turn told Oppenheimer, who  asked his student Snyder to work it out and this  led to
the first published paper on the subject.\ts{12} Today's string-theory origins of
noncommutativity are very similar to Peierls's application -- both rely on the presence of a
strong background field. 

When confronting the noncommutativity postulate~\refeq{e1}, it is natural to ask which
(infinitesimal) coordinate transformations 
\numeq{e2}{
\delta x^i = -f^i (x) 
}
leave \refeq{e1} unchanged. 
The answer is that the (infinitesimal) transformation vector function $f^i(x)$ must be
determined by a scalar through the expression\ts{13}
\numeq{e3}{
f^i (x) = \theta^{ij} \partial_j f(x) \ . 
}
Since $\partial_i f^i(x) = 0$,   these are recognized as volume-preserving transformations.
(They do not exhaust all volume preserving transformations, except in two dimensions. In
dimensions greater  two,
\refeq{e3} defines a subgroup of volume-preserving transforms that also leave $\theta^{ij}$
invariant.)

The volume-preserving transformations form the link between noncommuting coordinates
and fluid mechanics. Since the theory of fluid mechanics is not widely known outside the circle
of fluid mechanicians, let me put down some relevant facts.  There are two, physically
equivalent descriptions of fluid motion: One is the Lagrange formulation, wherein the fluid
elements are labeled, first by a discreet index~$n$: $\vec X_n(t)$ is the position as a function of
time of the $n$th fluid element.  Then one passes to a continuous labeling variable $n \to \vec
x: \vec X (t,\vec  x)$, and 
$\vec x$ may be taken to be the position of the fluid element at initial time $\vec X(0,\vec x) =
\vec x$. This is a comoving description. Because labels can be arbitrarily rearranged, without
affecting physical content, the continuum description is invariant against volume-preserving
transformations of $\vec x$, and in particular, it is invariant against the specific
volume-preserving transformations
\refeq{e3}, provided the fluid coordinate $\vec X$ transforms as a scalar:
\numeq{e4}{
\delta_f  \vec X  = f^i (\vec x) \frac\partial{\partial x^i} \vec X = \theta^{ij} \partial_i \vec
X\partial_j f\ . 
 }

The common invariance of Lagrange fluids and of noncommuting coordinates is a strong hint of
a connection between the two.

Formula \refeq{e4} will take  a very suggestive form when we rewrite it in terms of a bracket
defined for functions of $\vec x$ by 
\numeq{e5}{
\bigl\{ \mathcal O_1(\vec x), \mathcal O_2(\vec x)\bigr\} =\theta^{ij} \partial_i\mathcal
O_1(\vec x)
\partial_j\mathcal O_2(\vec x)\ .
} 
Note that with this bracket we have
\numeq{e6}{
\bigl\{x^i,  x^j\bigr\} = \theta^{ij}\ .
} 
So we can think of bracket relations as classical precursors of commutators for a
noncommutative field theory -- the latter obtained from the former by replacing brackets by
$-i$~times commutators, \`a la Dirac. More specifically, the noncommuting field theory that
emerges from the Lagrange fluid is a noncommuting U(1) gauge theory. 

This happens when the following steps are taken. We define the evolving portion of $\vec X$ by 
\numeq{e7}{
X^i (t,\vec x) = x^i + \theta^{ij} \hat A_j (t,\vec x)\ .
}
(It is assumed that $\theta^{ij}$ has an inverse.)
Then \refeq{e4} is equivalent to the suggestive expression
\numeq{e8}{
\delta_f \hat A_i = \partial_i f + \bigl\{\hat A_i, f\bigr\}\ .
}
When the bracket is replaced by $(-i)$ times the commutator, this is precisely the gauge
transformation for a noncommuting U(1) gauge potential $\hat A_i$. Moreover, the gauge field
$\hat F_{ij}$ emerges from the bracket of two Lagrange coordinates
\begin{gather}
\bigl\{ X^i, X^j\bigr\} = \theta^{ij} + \theta^{im} \theta^{jn} \hat F_{mn} \label{e9}\\
\hat F_{mn}  = \partial_m \hat A_n - \partial_n \hat A_m + \bigl\{\hat A_m, \hat  A_n\bigr\}\ . 
\label{e10}
\end{gather}
Again \refeq{e10} is recognized from the analogous formula
in noncommuting gauge theory.

What can one learn from the parallelism of the formalism for a Lagrange fluid and a
noncommuting gauge field? One result that has been obtained addresses the question of what
is  a gauge field's covariant response to a coordinate transformation. This question can be put
already for commuting, non-Abelian gauge fields, where conventionally the response is
given in terms of a Lie derivative $L_f$:
\begin{gather}
\delta_f x^\mu = - f^\mu(x) \label{e11}\\
\delta_f A_\mu = L_f A_\mu \equiv f^\alpha \partial_\alpha A_\mu + \partial_\mu f^\alpha
A_\alpha
 \ . 
\label{e12}
\end{gather}
But this implies
\numeq{e13}{
\delta_f F_{\mu\nu} = L_f F_{\mu\nu} \equiv f^\alpha \partial_\alpha F_{\mu\nu} + 
\partial_\mu f^\alpha F_{\alpha\nu} +  \partial_\nu f^\alpha F_{\mu\alpha}  
}
which is not covariant since the derivative in the first term on the right is not the covariant
one. The cure in this, commuting, situation has been given some time ago:\ts{14} Observe that
\refeq{e12} may be equivalently presented as 
\numeq{e14}{
\begin{aligned}
\delta_f A_\mu = L_f A_\mu &= f^\alpha \bigl(
\partial_\alpha A_\mu -\partial_\mu A_\alpha - i [A_\alpha, A_\mu]
\bigr) \\
&\qquad{}+ f^\alpha \partial_\mu A_\alpha  - i [  A_\mu, f^\alpha A_\alpha] + 
\partial_\mu f^\alpha A_\alpha\\
 &= f^\alpha F_{\alpha\mu} +  D_\mu (f^\alpha A_\alpha)\ .
\end{aligned}
}
Thus, if the coordinate transformation  generated by $f^\alpha$ is supplemented by a  gauge
transformation generated by $-f^\alpha A_\alpha$, the result is a gauge covariant coordinate
transformation
\numeq{e15}{
\delta'_f A_\mu = f^\alpha F_{\alpha\mu}
}
and the modified response of $F_{\mu\nu}$ involves the gauge-covariant Lie derivative $L'_f$:
\numeq{e16}{
\delta'_f F_{\mu\nu} = L'_f F_{\mu\nu} \equiv f^\alpha D_\alpha F_{\mu\nu} + 
\partial_\mu  f^\alpha F_{\alpha\nu} +  \partial_\nu f^\alpha  F_{\mu\alpha} \ .
}

In the noncommuting situation, loss of covariance in the ordinary Lie derivative is even
greater, because in general the coordinate transformation functions $f^\alpha$ do not
commute with the fields $A_\mu, F_{\mu\nu}$; moreover, multiplication of $x$-dependent
quantities is not a covariant operation. All these issues can be addressed and resolved by
considering them in the fluid mechanical context, at least, for volume-preserving
diffeomorphisms. The analysis is technical and I refer you to the published papers.\ts{13,15}

Instead, I shall discuss the Seiberg-Witten map,\ts{16} which can be made very
transparent by the fluid analogy.  The Seiberg-Witten map replaces the noncommuting vector
potential $\hat A_\mu$  by a nonlocal function of a commuting potential $a_\mu$ and
of~$\theta$;  i.e., the former is viewed as a function of the latter.  The relationship between the
two follows from the requirement of stability against gauge transformations: a noncommuting
gauge transformation  of the noncommuting gauge potential should be equivalent to a
commuting gauge transformation on the commuting vector potential on which the
noncommuting potential depends. Moreover, when the action and the equations of motion of
the noncommuting theory are transformed into commuting variables, the dynamical content
is preserved: the physics described by noncommuting variables is equivalently described by
the commuting variables, albeit in a complicated, nonlocal fashion. 

  The Seiberg-Witten map is intrinsically interesting in the unexpected equivalence that it
establishes. Moreover, it is practically useful for the following reason. It is difficult to extract
gauge invariant content from a noncommuting gauge theory because quantities constructed
locally from $\hat F_{\mu\nu}$ are not gauge invariant; to achieve gauge invariance, one must
integrate over space-time. Yet for physical analysis one wants local quantities: profiles of
propagating waves, etc.  Such local quantities can be extracted in a gauge invariant manner
from the physically equivalent, Seiberg-Witten mapped commutative gauge theory.\ts{17} 

Let me now use the fluid analogy to obtain an explicit formula for the Seiberg-Witten map;
actually, we shall present the inverse map, expressing commuting fields in terms of
noncommuting ones. For our development we must refer to a second, alternative formulation
of fluid mechanics, the so-called Euler formulation. This is not a comoving description, rather
the experimenter observes the fluid density~$\rho$ and velocity~$\vec v$ at given point in
space-time $(t,\vec r)$. The current is $\rho \vec v$ and satisfies with $\rho$ a continuity
equation
\numeq{e17}{
\frac\partial{\partial t} \rho + \grad \cdot (\rho\vec v) = 0\ . 
}
The theory is completed by  positing an ``Euler equation'' for $\partial\vec v/\partial t$, but
we shall not need this here. 

Of interest to us is the relation between the Lagrange description and the Euler description.
This is given by the formulas
\begin{subequations}\label{e18}
\begin{gather}
\rho(t,\vec r) = \int \rd x \delta\bigl(\vec X(t,\vec x) -\vec r\bigr)\label{e18a}\\
\rho(t,\vec r)\vec v(t,\vec r) \equiv \vec j(t,\vec r) = 
    \int \rd x \frac\partial{\partial t} \vec X(t,\vec x) \delta \bigl(\vec X(t,\vec x) -\vec
r\bigr)\ . \label{e18b}
\end{gather}
\end{subequations}
(The integration and the $\delta$-function carry the dimensionality of space.) Observe that
the continuity equation \refeq{e17} follows from the definitions \refeq{e18}, which can be
summarized by 
\begin{gather}
j^\mu(t,\vec r) = \int \rd r \frac{\partial}{\partial t}  X^\mu \delta (\vec X - \vec
r)\label{e19}\\ X^0 = t\notag\\
\partial_\mu  j^\mu = 0\ . \label{e20}
\end{gather}

The (inverse) Seiberg-Witten map, for the case of two spatial dimensions, can be extracted
from \refeq{e19}, \refeq{e20}.\ts{13} (The argument can be generalized to arbitrary
dimensions, but there it is more complicated.\ts{13}) Observe that the right side of \refeq{e19}
depends on
$\hat A$ through
$\vec X$ [see
\refeq{e7}]. It is easy to check that the integral \refeq{e19} is invariant under the
transformations \refeq{e4}; equivalently viewed as a function of $\hat A$, it is gauge invariant  
[see \refeq{e8}]. Owing to the conservation of $j^\mu$   [see \refeq{e20}], its dual 
$\eps_{\alpha\beta\mu} j^\mu$ satisfies a conventional, commuting Bianchi identity, and
therefore can be written as the curl of an Abelian vector potential~$a_\alpha$, apart from
proportionality  and additive constants: 
\numeq{e21}{
\begin{gathered}
\partial_\alpha a_\beta - \partial _\beta a_\alpha + \text{constant} 
\propto \eps_{\alpha\beta\mu} \int \rd x \frac\partial{\partial t} X^\mu \delta(\vec X- \vec
r)\\
\partial_i a_j - \partial _j a_i + \text{constant} 
\propto \eps_{ij} \int \rd x   \delta(\vec X- \vec r) = \eps_{ij} \rho\ .
\end{gathered}
 }
This is the (inverse) Seiberg-Witten map, relating the~$a$ to~$\hat A$. 

Thus far operator noncommutativity has not been taken into account. To do so, we must
provide an ordering for the $\delta$-function depending on the operator $X^i = x^i + \theta^{ij}
\hat A_j$.  This we do with the Weyl prescription by Fourier transforming. The  final operator
version of equation~\refeq{e21}, restricted to the two-dimensional spatial components, reads
\numeq{e22}{
\int \rd r e^{i\vec k\cdot\vec r} (\partial_i a_j - \partial_j a_i) = 
-\eps^{ij} \Bigl[ \int \rd x  e^{i\vec k\cdot\vec X} - (2\pi)^2\delta(\vec k)   \Bigr]\ . 
}
Here the additive and proportionality constants are determined by requiring agreement for
weak noncommuting fields, and the integral on the right is interpreted as a trace over the
operators.

Formula~\refeq{e22} has previously appeared  in a direct analysis of the Seiberg-Witten
relation.\ts{18} Here we recognize it as the (quantized) expression relating Lagrange and Euler
formulations for fluid mechanics. 
\nopagebreak

I think \Loch\ would have liked this.

\small
\def\Journal#1#2#3#4{{\em #1} {\bf #2}, #3 (#4)}
\def\add#1#2#3{{\bf #1}, #2 (#3)}
\def\Book#1#2#3#4{{\em #1}  (#2, #3 #4)}
\def\Bookeds#1#2#3#4#5{{\em #1}, #2  (#3, #4 #5)}
% \Journal{}{}{}{}
% \Book{}{}{}{}

\end{document}